\documentstyle[preprint,aps,epsfig]{revtex}
\catcode`\@=11

\newcommand{\beq}{\begin{equation}}
\newcommand{\eeq}{\end{equation}}

\newcommand{\be}{\begin{equation}}
\newcommand{\ee}{\end{equation}}
\newcommand{\bea}{\begin{eqnarray}}
\newcommand{\eea}{\end{eqnarray}}

\newcommand{\ba}{\begin{array}}   %
\newcommand{\ea}{\end{array}}     %
\begin{document}
\baselineskip=17pt

\gdef\journal#1, #2, #3, 1#4#5#6{{#1~}{\bf #2}, #3 (1#4#5#6)}
\gdef\ibid#1, #2, 1#3#4#5{{\bf #1} (1#3#4#5) #2}

\title{
Projection on higher Landau levels and non Commutative Geometry}

\author{Nicolas Macris and St\'ephane Ouvry }

\date{December 19, 2001}

\address{Institut de Physique Th\'eorique, \'Ecole Polytechnique
F\'ed\'erale de Lausanne, CH-1015, Lausanne, Suisse}
\address{Laboratoire de Physique Th\'eorique et Mod\`eles
Statistiques, UMR 8626 CNRS, B\^at. 100, Universit\'e Paris-Sud, 91405 Orsay, France}

\maketitle
{\begin{abstract}
The projection of a two dimensional planar system on the  higher
Landau levels
of an external magnetic field is formulated in the language of the 
non commutative plane and leads to a  new class of
 star products.

\end{abstract}}
\narrowtext

\newpage
{\bf Introduction:} Consider a two dimensional Hamiltonian of a particle in
  a scalar potential $V(z,\bar z)$ coupled to a magnetic field B, in the
  symmetric gauge ($e=\hbar=m=1$)
\be\label{1} H(z)=-2\partial\bar \partial+\omega_c(\bar z\bar\partial
-z\partial)+{1\over 2}\omega_c^2z\bar z+V(z,\bar z)
\ee
We assume without any loss of generality that $B\ge 0$ and denote by
$\omega_c=+B/2$
 half the cyclotron frequency.
 It is well known that by projecting (\ref{1})  on the lowest Landau level (LLL)
 spanned by the states
\be\label{2}\psi(z)=f(z)e^{-{\omega_c\over 2}z\bar z}\ee
where $f(z)$ is analytic,
one obtains an eigenvalue equation
\be\label{3}:{V}(z,{1\over \omega_c}\partial_z):
f(z)=(E-\omega_c)f(z)\ee
The normal ordering $:\quad :$ means that the
differential operator ${1\over \omega_c}\partial$ is put on the left of $z$.
Equation (\ref{3}) is a reformulation of the Peierls substitution \cite{Pei}
(which does not specify the correct ordering in general) and as such has
been derived in \cite{GJ} (which specifies the correct ordering).

Clearly,
the usual two dimensional plane has been traded for a non commutative space
\be\label{4} [{1\over \omega_c}\partial_z,z]={1\over \omega_c}\ee
In view of the above commutation relation, the non commutative space can be
interpreted as the phase space corresponding to a one dimensional space.
However this interpretation may not be entirely satisfactory because $z$ is
a complex coordinate. There is another natural interpretation in terms of a
non commutative space which is a two dimensional plane with non commuting
"real" coordinates $(X,Y)$. It is well known that the algebra of operators
depending on the two non commutative coordinates $(X,Y)$ is equivalent to a
deformation of the classical algebra of functions on the usual commutative
plane with coordinates $(x,y)$. This deformation is defined through a non
commutative star product. Although its construction is well known we will
review it
below in a self contained way, giving, as a by product, a more systematic
and more simple derivation of (\ref{3}).

The main point of this letter is to show that non commutative geometry is by
no means specific to the LLL projection but can be obtained as well by
projecting the two dimensional system on any given higher Landau level. As an
illustration the case of the first  Landau level (1LL) will be
analyzed in detail. We will generalise the Peierls substitution to the 1LL
and reformulate it
in a non commutative plane language. We will find that although the canonical
commutation
relation between the non commutative coordinates will be the same as for the LLL,
a new non commutative star product will appear to be naturally connected
with the 1LL. We
will then give the general expression for the star product associated to
any given
Landau level, thereby defining a new class of star product \cite{Kon}.

It might be objected that projecting a system on a higher Landau level
is counter intuitive: usually the LLL projection is regarded as physically
justified when the cyclotron gap $\hbar\omega_c$ is sufficiently large
-thus the LLL projection is associated
to a strong magnetic field-
compared to the temperature and/or to the potential so that the excited
states above the LLL can be ignored. Clearly such an interpretation becomes
meaningless in a higher Landau level. However,
restricting the two dimensional Hilbert space to a
given Landau level subspace is a well defined
mathematical procedure, and will be considered as such in what follows.

Before starting let us  remind that
the  Landau   spectrum is made of
degenerate Landau levels $ (2n+1)\omega_c, n\ge 0$
with in the $n$th  Landau level
eigenstates labelled by the radial/orbital quantum numbers $n, l\ge 0$
(analytic)
and  $n+l, -n\le l<0$  (anti-analytic).
There is, in a given landau level,  an infinite number of analytic
eigenstates
\be\label{5}
<z,\bar z|n,l>=z^l L_n^l(\omega_cz\bar z)
e^{-{1\over 2}\omega_cz\bar z}\quad l\ge 0
\ee
and a finite number of anti-analytic eigenstates
\be\label{5bis}
<z,\bar z|n+l, l>= \bar z^{-l}
L_{n+l}^{-l}(\omega_cz\bar z)e^{-{1\over 2}\omega_cz\bar z} \quad -n\le l<0
\ee

\vspace{1.5cm}

{\bf Projection on the  LLL: }
In the lowest Landau level, $n=0$, $l\ge 0$,
 the   eigenstates are analytic
\be \label{6}
<z,\bar z|0,l>=({{\omega_t^{\ell+1}\over \pi \ell!}})^{{1\over 2}}z^{\ell}
e^{-{1\over 2}\omega_tz\bar z},\quad
\ell\ge 0\ee
Consider the projector on the
LLL Hilbert space
$ P_o=\sum_{l=0}^{\infty}|0,l><0,l|$
\be \label{8} <z,\bar z|P_o|z',\bar z'>={\omega_c\over \pi}
e^{-{\omega_c\over 2}(z\bar z+z'\bar z'-2z\bar z')}\ee
A state  of the LLL
$|\psi>=\sum_{l=0}^{\infty}a_l|0,l>$,  is analytic up to
the Landau
gaussian factor
\be\label{10} <z|\psi>=f(z)e^{-{\omega_c\over 2}z\bar z}\ee
with
\be \label{11} f(z)=\sum_{l=0}^{\infty}a_l z^l\ee
One can check that $|\psi>$ satisfies  $ P_o|\psi>=|\psi>$.

Then project the Hamiltonian (\ref{1}) on the LLL
\be \label{12} <z,\bar z|P_oHP_o|\psi>=<z,\bar z|P_oH|\psi>=\int dz'd\bar z'{\omega_c\over \pi}
e^{-{\omega_c\over 2}(z\bar z+z'\bar z'-2z\bar z')}H(z')f(z')e^{-{\omega_c\over 2}z'\bar z'}\ee
Using the Bargman identity
\be
\label{barg}{\omega_c\over \pi}\int dz' d\bar z'
e^{-{\omega_c}(z'\bar z'-z\bar z')}h(z')=h(z)
\ee
it is not difficult to see that the eigenvalue equation
\be \label{13} <z,\bar z|P_oHP_o|\psi>=E<z,\bar z|\psi>\ee
or
\be\label{14}\int dz'd\bar z'{\omega_c\over \pi}
e^{-{\omega_c}(z'\bar z'-z\bar z')}
\left( \omega_c+ V(z',\bar z')\right)f(z')=Ef(z)\ee
can be transformed into a differential equation which is precisely
the Peierls substitution equation
(\ref{3}).

\vspace{1.5cm}

{\bf Projection on the 1LL:}
Consider now
the
projector on the
1LL Hilbert space
$ P_1=\sum_{l=0}^{\infty}|1,l><1,l|+|0,-1><0,-1|$
\be \label{18}<z,\bar z|P_1|z',\bar z'>=<z,\bar z|P_0|z',\bar z'>[1-\omega_c(z'-z)(\bar
z'-\bar z)
]\ee

Using $L_1^l(\omega_cz\bar z)=l+1-\omega_c z\bar z$
one obtains that  a state  of the 1LL
 $|\psi>=\sum_{l=0}^{\infty}a_l|1,l>+a_{-1}|0,-1>$ is of the form.

\be\label{20}<z,\bar z|\psi>=\left(f(z)+\bar zg(z)\right)e^{-{\omega_c\over 2}z\bar z}\ee
with $f(z)$ and $g(z)$ analytic,
\be\label{21} f(z)=-{1\over \omega_c}\partial g(z)\ee
and
\be \label{22} g(z)=-\omega_c\sum_{l=0}^{\infty}a_l z^{l+1}+a_{-1}\ee

Another way to recover this result is to impose equivalently that
$P_1|\psi>=|\psi>$, i.e.
\be\label{23} \int dz'd\bar z'<z,\bar z|P_1|z',\bar z'><z',\bar z'|\psi>=<z,\bar
z|\psi>\ee
 One first infers that necessarily $<z,\bar z|\psi>=\left(f(z)+\bar
zg(z)\right)e^{-{\omega_c\over 2}z\bar z}$.
Equation (\ref{23}) then becomes
\be\label{24} \int dz'd\bar z'{\omega_c\over \pi}
e^{-{\omega_c}(z'\bar z'-z\bar z')}[1-\omega_c(z'-z)(\bar
z'-\bar z)
]\left(f(z')+\bar
z'g(z')\right)=f(z)+\bar
zg(z)\ee
and finally
\be \label{25}-{1\over \omega_c}\partial g(z) +\bar z
g(z)
=f(z)+\bar zg(z)\ee
The relation (\ref{21}) directly follows.

Now we project the Hamiltonian (\ref{1}) on the 1LL,
\be\label{26} <z,\bar z|P_1HP_1|\psi>=\int dz'd\bar z'
<z,\bar z|P_1|z',\bar z'><z',\bar z'|H|\psi>\ee
The eigenvalue equation
\be <z,\bar z|P_1HP_1|\psi>=E<z,\bar z|\psi>\ee
becomes
\bea\label{27} \int dz'd\bar z'{\omega_c\over \pi}
e^{-{\omega_c}(z'\bar z'-z\bar z')}[1-\omega_c(z'-z)(\bar
z'-\bar z)
]
&V&(z',\bar z')(f(z')+\bar z'g(z'))\nonumber\\&=&(E-3\omega_c)(f(z)+\bar z g(z))\eea
Using again (\ref{barg})
one  transforms (\ref{27}) into the differential equation
\bea\label{28}-{1\over \omega_c}\partial\left(:V
+{1\over \omega_c}\partial\bar \partial V:g(z)\right)
&+&\bar z :V+{1\over \omega_c}\partial\bar \partial V
:g(z)
\nonumber\\&=&(E-3\omega_c)(f(z)+\bar z g(z))
\eea
where  it is again  understood that
 $\bar z$ is replaced by ${1\over \omega_c}\partial$
both in $V$ and in $\partial\bar \partial V$, and then the
normal ordering is taken.

As an example, consider $V(z,\bar z)=\omega_c^2z\bar z/2$.
We obtain from (\ref{28})
\bea
{\omega_c\over 2}(3+z\partial)g&=&(E-3\omega_c)g(z)\nonumber\\
\label{29} {\omega_c\over 2}(4+z\partial)f&=&(E-3\omega_c)f(z)\eea
It is easy to see that
the couple of equations lead to  the  spectrum
$E-3\omega_c=\omega_c(3+l)/2, l\ge 0$.
For a general $V(z,\bar z)$ this conclusion still holds true:
the 1LL projection
induces an eigenvalue equation
\be\label{30} :V(z,\frac{1}{\omega_c}\partial)+\frac{1}{\omega_c}
\bar\partial\partial V(z,\frac{1}{\omega_c}\partial):
g(z)=(E-3\omega_c)g(z)\ee
where $g(z)$ is analytic.
The equation for $f(z)$ is obtained by differentiating (\ref{30})
with respect to $z$. Equation (\ref{30}) is, in the 1LL, the analogous of
the LLL Peierls
substitution equation (\ref{3}).

\vspace{1.5cm}

{\bf Non commutative plane and $\star$ products:}
It is well-known that the LLL quantum mechanics
can be equivalently reformulated in a non commutative geometry setting.
It is tempting to generalize this construction to  higher Landau levels, with as a result a class of
non commutative $*_n$ products associated to the
higher Landau level projection.
Let us first start by a reminder of the LLL non commutative
formulation.

\vspace{1.5cm}

\noindent {\bf Lowest Landau Level:}
Consider the  non-commutative guiding center
coordinates
\be\label{31}
X=\frac{1}{2}(z+\frac{1}{\omega_c}\partial ),
Y=\frac{1}{2i}(z-\frac{1}{\omega_c}\partial )
\ee
They satisfy the commutation relation
\be
\label{32}
[X,Y]=\frac{1}{2i\omega_c}
\ee
We have seen in (\ref{3}) that  the LLL
 operator associated to the classical potential $V(x,y)$ is
\be\label{33}
:V(z, \frac{1}{\omega_c}\partial z):
\ee
Introducing the Fourier transform of the classical potential
\be\label{34}
V(x,y)=\int dk dl  e^{i(kx+ly)}\tilde V(k, l)=
\int dk dl  e^{i(\frac{k}{2}(z+\bar z)+\frac{l}{2i}(z-\bar z))}\tilde V(k, l)
\ee
and using (\ref{31}) the operator (\ref{33}) can be written as
\be\label{35}
\hat V^{(0)}(X, Y)=\int dk dl :e^{ikX+ilY}: \tilde V(k, l)
\ee
inducing a mapping between a classical potential and an operator.
Note that (\ref{35}) is different from the Weyl ordering which is often used
in the field theory context \cite{H} and corresponds to drop the normal order
double colons in (\ref{35}). Note also for further use that, thanks to the
identity
\be\label{48}
:e^{ikX+ilY}:=e^{-\frac{1}{8\omega_c}(k^2+l^2)} e^{ikX+ilY}
\ee
one has
\be\label{36}
\hat V^{(0)}(X,Y)=\int dk dl e^{-\frac{1}{8\omega_c}(k^2+l^2)} e^{ikX+ilY} \tilde V(k, l)
\ee
Equation (\ref{35}) induces a non commutative product $f\star_0g$ between
 any classical functions $f(x,y)$ and $g(x,y)$ such that
\be\label{37}
\hat f^{(0)}(X,Y) \hat g^{(0)}(X,Y) = \widehat{(f\star_0g)}^{(0)} (X,Y)
\ee
A straightforward computation leads to
\be\label{38}
(f\star_0g)(x,y)= e^{-\frac{1}{4\omega_c}(\partial_x^\prime+i\partial_y^\prime)
(\partial_x-i\partial_y)}f(x,y)g(x^\prime,y^\prime)|_{x=x^\prime, y=y^\prime}
\ee
and
\be\label{39}
\widetilde{(f\star_0g)}(k,l)=\int dk^\prime dl^\prime \tilde f(k-k^\prime, l-l^\prime)\tilde g(k^\prime, l^\prime) e^{\frac{i}{4\omega_t}(kl^\prime-lk^\prime)}
e^{-\frac{1}{4\omega_c}(k^{\prime2}+l^{\prime2}-kk^{\prime}-ll^{\prime})}
\ee
The Fourier transform (\ref{39}) of $f\star_0g$  is a "deformation"
of the usual convolution product of the Fourier transforms of $f$ and $g$. The canonical commutation relation may be expressed as
\be\label{40}
x\star_0y -y\star_0x=\frac{1}{2i\omega_c}
\ee
In terms of the non commutative coordinates $(X,Y)$
the eigenvalue equation (\ref{3})
acting on $f(z)$ rewrites
\be\label{41}
\hat V^{(0)}(X,Y)f(X+iY) =(E-\omega_c)f(X+iY)\ee
Looking at $f(X+iY)\equiv \hat f^{(0)}$ as an operator and using (34) one finds that
(\ref{41}) is equivalent to
\be\label{nc1}
\widehat{(V\star_0 f)}^{(0)} (X,Y)=(E-\omega_c)f(X+iY)
\ee
In other words the eigenvalues of (\ref{3})  and of the operator
equation (\ref{nc1}) are identical.
Note finally that the analyticity of $f$ implies
$V\star_0 f=Vf$, and therefore
$\widehat{(V f)}^{(0)} (X,Y)=(E-\omega_c)f(X+iY)$
\footnote{Of course it does not
imply that $Vf(x+iy)=(E-\omega_c)f(x+iy)$.}.

\vspace{1.5cm}
\noindent {\bf First Landau Level:}
As seen in (\ref{30}), the 1LL  operator associated to the
potential $V(x,y)$ is
\be\label{45}
:V(z,\frac{1}{\omega_c}\partial)+\frac{1}{\omega_c}
\bar\partial\partial V(z,\frac{1}{\omega_c}\partial):
\ee
This can be reexpressed as
\be
\label{46}
\hat V^{(1)}(X,Y)=\int dk dl :e^{ikX+ilY}: (1-
\frac{1}{4\omega_c}(k^2+l^2))\tilde V(k,l)
\ee
Using the commutation relation (\ref{32}) one can check that
\be
\label{47}
(k^2+l^2)e^{ikX+ilY}=[X,[X,e^{ikX+ilY}]+[Y,[Y,e^{ikX+ilY}]
\ee
Thus from (\ref{48}) and  (\ref{47}) we find
\be
\label{49}
\hat V^{(1)}(X,Y)=\hat V^{(0)}(X,Y)+\frac{1}{4\omega_c}\hat\Delta\hat V^{(0)}(X,Y)
\ee
where in (\ref{49})
\be\label{50}
\hat \Delta=[X,[.,X]+[Y,[.,Y]
\ee
and
\be
\label{51}
\widehat{(\Delta V)}^{(0)}(X,Y)=\hat\Delta \hat V^{(0)}(X,Y)
\ee
is understood.
Equation (\ref{47}) suggests that the operator (\ref{50}) is the non
commutative version of the Laplacian \cite{Gross}.

The 1LL mapping (\ref{46}) between a classical function and
an operator induces a new
star product $\star_1$ such  that for any classical functions $f(x,y)$ and $g(x,y)$
\be\label{52}
\hat f^{(1)}(X,Y)\hat g^{(1)}(X,Y)=\widehat{(f\star_1 g)}^{(1)}(X,Y)
\ee
A differential expression for the $\star_1$ product can easily be found
 using  (\ref{51}). Indeed,
\bea
\hat f^{(1)}\hat g^{(1)}&=&(\hat f^{(0)}+\frac{1}{4\omega_c}\hat\Delta \hat f^{(0)})
(\hat g^{(0)}+\frac{1}{4\omega_c}\hat\Delta \hat g^{(0)})\nonumber\\
&=&(\hat f^{(0)}+\frac{1}{4\omega_c}\widehat {\Delta f}^{(0)})
(\hat g^{(0)}+\frac{1}{4\omega_c}\widehat{\Delta g}^{(0)})\label{53}
\eea
so that
\be\label{54}
f\star_1 g=(1+\frac{1}{4\omega_c}\Delta)^{-1}\biggl((f+\frac{1}{4\omega_c}\Delta f)\star_0(g+\frac{1}{4\omega_c}\Delta g)\biggr)
\ee
Thus
\be\label{55}
f\star_1 g=(1+\frac{1}{4\omega_c}\Delta)^{-1}\biggl(e^{-\frac{1}{4\omega_c}(\partial_x-i\partial_y)(\partial_x'+
i\partial_y')}(1+\frac{1}{4\omega_c}\Delta)(1+\frac{1}{4\omega_c}\Delta')
f(x,y)g(x',y')|_{x=x', y=y'}\biggr)
\ee
Note that it follows from (\ref{55}) that the canonical commutation relation
$x\star_1y -y\star_1x=\frac{1}{2i\omega_c}$
is
identical to (\ref{40}), i.e. it is the same as in  the LLL.

In terms of the non commutative coordinates $X,Y$ the eigenvalue equation
(\ref{30})
becomes
\be\label{nc2}
\hat V^{(1)}(X,Y) g(X+iY)=(E-3\omega_c)g(X+iY)
\ee
or equivalently
\be\label{nc3}
\widehat{(V\star_1 g)}^{(1)}(X,Y)=(E-3\omega_c)g(X+iY)
\ee
Note that since $g(z)$ is analytic
$V\star_1 g=
(1+\frac{1}{4\omega_c}\Delta)^{-1}((V+\frac{1}{4\omega_c}\Delta V)g)$.

\vspace{1.5cm}

\noindent{\bf Higher Landau level:} Let us apply
the  same procedure than in the LLL and the 1LL to the the  $n$-th Landau level (nLL).
An eigenstate is of the form
\be\label{wf}
<z,\bar z|\psi>=(f_0(z)+\bar zf_1(z)+...+\bar z^n f_n(z))
e^{-\frac{\omega_c}{2}z\bar z}
\ee
with $f_n(z)$ analytic and
\be\label{constr}
f_i(z)=\omega_c^{(i-n)}(-1)^{(n-i})\frac{n!}{i!(n-i)!}\frac{\partial^{n-i}}
{\partial z^{n-i}} f_n(z)
\ee
Projecting the Hamiltonian (\ref{1}) on the nLL implies that
$f_n(z)$ satisfies to the eigenvalue equation
\be\label{gjg}
:\sum_{i=0}^n{1\over \omega_c^i}
\frac{n!}{i!^2(n-i)!}\frac{\partial^{2i}}{\partial z^i\partial\bar
 z^i}V:f_n(z)=(E-(2n+1)\omega_c)f_n(z)
\ee
which can be viewed as the generalisation of the LLL Peierls substitution
equation (\ref{3}) to the nLL
(details of the derivation of (\ref{wf},\ref{gjg}) can be found in
\cite{Nina}).

In a non commutative plane formulation, the potential $V(x,y)$ is
replaced by the nLL operator
\be\label{op}
\hat V^{(n)}(X,Y)=\int dk dl :e^{ikX+ilY}:\tilde V(k,l)\sum_{i=0}^n
(-{1\over 4\omega_c})^{i}
\frac{n!}{i!^2(n-i)!}(k^2+l^2)^i
\ee
with also
\be \label{form}
\hat V^{(n)}(X,Y)=\sum_{i=0}^n
({1\over 4\omega_c})^{i}
\frac{n!}{i!^2(n-i)!}(\hat \Delta)^i \hat V^{(0)}(X,Y)
\ee
This induces a  $\star_n$ product between classical functions
such that $\hat f^{(n)}\hat g^{(n)}=(\widehat{ f\star_n g})^{(n)}$:
namely one has
\be\label{stn} f\star_n g =(D_{x,y}^{(n)})^{-1}\biggl(
e^{-\frac{1}{4\omega_c}(\partial_x-i\partial_y)(\partial_x'+
i\partial_y')}D_{x,y}^{(n)}D_{x^\prime,y^\prime}^{(n)}
f(x,y)g(x',y')|_{x=x', y=y'}\biggr)
\ee
where $D_{x,y}^{(n)}=\sum_{i=0}^n
({1\over 4\omega_c})^{i}
\frac{n!}{i!^2(n-i)!}(\Delta)^i$.
Note that the canonical commutation relation  associated
to $\star_n$ again narrows down to (\ref{40}), i.e. to the LLL
situation.

\vskip 1.5cm

\noindent{\bf Hilbert space:} Equation (\ref{op}) gives a natural
specification of nLL operators $\hat f^{(n)}(X,Y)$ associated to classical functions
$f(x,y)$. One can also construct a  Hilbert space on which these
operators act and which turns out to be equivalent to the Bargmann space of
analytic functions with the scalar product $<f|g>=(\omega_c/\pi)\int dzd\bar
z e^{-\omega_c z\bar z}\bar f g$.

Let us define the creation-annihilation operators $A^\dagger=\omega_c^{1/2}
(X+iY)$ and $A=\omega_c^{1/2}(X-iY)$
and the corresponding  Hilbert space  spanned by the
vectors $A^{\dagger n}|0>$ acting on
the  vacuum $|0>$ with $A|0>=0$.
Since
$[A,A^\dagger]=1$ we have
\be\label{vac}
<0|e^{i(kX+lY)}|0>=e^{-\frac{1}{2}<0|(kX+lY)^2|0>}
=e^{-\frac{1}{8\omega_c}(k^2+l^2)}
\ee
If we now consider  in the Hilbert space two states $\hat
f^{(n)}|0>$ and $\hat g^{(n)}|0>$ their scalar product becomes
\bea\label{sc}
<0|\hat f^{(n)\dagger} \hat g^{(n)}|0>&=&
<0|\widehat{(\bar f\star_n g)}^{(n)}|0>\nonumber\\&=&\int dkdle^{-\frac{1}{4\omega_c}(k^2+l^2)}
\widetilde{(\bar f\star_n g)}(k,l)
\sum_{i=0}^n
(-{1\over 4\omega_c})^{i}
\frac{n!}{i!^2(n-i)!}(k^2+l^2)^i
\nonumber\\
&=& {\omega_c\over \pi}\int dxdye^{-\omega_c(x^2+y^2)}
 D_{x,y}^{(n)}(\bar f\star_n g)(x,y)
\eea
The scalar product (\ref{sc}) is equivalent to the scalar product of the
Bargmann space. Indeed for any  classical function $f(x,y)$, one can always
define
a polynomial (analytic)
function of $A^\dagger$ such that  $\hat f^{(n)}|0> =
p_f(A^\dagger)|0>$. The analyticity then implies that 
$p_f\star_n p_g=(D_{x,y}^{(n)})^{-1}(\bar p_f p_g)$ so that
\be\label{bb}
<0|\hat f^{(n)\dagger} \hat g^{(n)}|0>=
{\omega_c\over \pi}\int dxdye^{-\omega_c(x^2+y^2)}
\bar p_f(x-iy) p_g(x+iy)
\ee
Equations (\ref{nc1}) and (\ref{nc3}) can be viewed as  bona fide
eigengenvalue equations on
the Hilbert space spanned by $(A^\dagger)^n|0>$,
\be\label{eig}
\hat V^{(n)}(X,Y) f(X+iY)|0>=(E-(2n+1)) f(X+iY)|0>
\ee
Note finally that in (\ref{sc})  
 scalar products have been expressed in the operator language or 
equivalently in
terms of  star products, leading to non trivial
identities\footnote{The scalar product
\be\label{pos}
<0|\hat f^{(n)\dagger} \hat f^{(n)}|0>
={\omega_c\over \pi}\int dxdye^{-\omega_c(x^2+y^2)}
D_{x,y}^{(n)}(\bar f\star_n f)(x,y)
\ee
must be non negative for any  $f(x,y)$ although
\be\label{pos2}
D_{x,y}^{(n)}(\bar f\star_n f)(x,y)=\sum_{j=0}^\infty \frac{(-1)^j}{j!} (4\omega_c)^{-j}
|(\partial_x+i\partial_y)^jD_{x,y}^{(n)}f(x,y)|^2
\ee
is an alternating sum. As an illustration consider in the LLL ($n=0$)
  $f(x,y)=(x+iy)^k(x-iy)^l$. Then $\hat f^{(0)}(X,Y)
=(X-iY)^l(X+iY)^k$ so that for $l>k$ we have $\hat f^{(0)}|0>=0$. 
This is satisfied if, evaluating the right hand side of (\ref{pos}),
the combinatorial identity
\be\label{comb}
\sum_{j=0}^l(-1)^j{(j+k)(j+k-1)...(j+1)\over(l-j)!j!}=0, \qquad l>k\geq 0
\ee
is verified.
An independent check can be made using the binomial expansion
of $\frac{d^k}{dx^k}(1-x)^l|_{x=1}=0$ for $l>k\geq 0$.
}.

{\bf Conclusion:}
One has obtained
the class of star products $\star_n$ which generalize to the nth Landau level the standard
$\star_0$
product in the LLL. A non commutative space can be thus defined each time
the two dimensionnal Hilbert space is projected on a given Landau level.
Accordingly, a nLL Peierls subtsitution equation is obtained which
generalizes the standard LLL Peierls substitution equation.
In each nLL subspace the space is non commutative with a canonical commutation
relation $x\star_ny -y\star_nx=\frac{1}{2i\omega_c}$. However, by considering altogether all
the nLL projections one should recover the full Landau spectrum
and therefore the
 commutative two dimensionnal  space.



\vskip 0.5cm 


\begin{references}


\bibitem{Pei} R. Peierls, Z. Phys. {\bf 80}, 763 (1933)

\bibitem{GJ} S. Girvin and T. Jachs,
Phys. Rev. D{\bf 29}, 5617 (1984);
G. Dunne and R.
Jackiw, hep-th/9204057; R.Jackiw, hep-th/0110057

\bibitem{Kon} for general $\star$ product see M. Kontsevich, ``Deformation Quantization of Poisson
Manifolds, I'', q-alg/9709040

\bibitem{H} for a recent review see for example R. J. Szabo, hep-th/0109162

\bibitem{Gross} for non commutative Laplacian see D. J.
Gross and N. A. Nekrasov, hep-th/0005204

\bibitem{Nina}
 N. Rohringer, Lausanne report (2000)
\end{references}
\end{document}